# Towards Subject and Diagnostic Identifiability in the Alzheimer's Disease Spectrum based on Functional Connectomes

Diana O. Svaldi*[1][0000-0001-6265-1931], Joaquín Goñi*[1,2][0000-0003-3705-1955], Apoorva Bharthur Sanjay[1], Enrico Amico[2], Shannon L. Risacher[1], John D. West[1], Mario Dzemidzic[1], Andrew Saykin[1], and Liana Apostolova[1]

[1] Indiana University School of Medicine, Indianapolis IN 46202, USA
[2] Purdue University, Lafayette IN 47907, USA
`dosvaldi@iu.edu, jgonicor@purdue.edu`

**Abstract.** Alzheimer's disease (AD) is the only major cause of mortality in the world without an effective disease modifying treatment. Evidence supporting the so called "disconnection hypothesis" suggests that functional connectivity biomarkers may have clinical potential for early detection of AD. However, known issues with low test-retest reliability and signal to noise in functional connectivity may prevent accuracy and subsequent predictive capacity. We validate the utility of a novel principal component based diagnostic identifiability framework to increase separation in functional connectivity across the Alzheimer's spectrum by identifying and reconstructing FC using only AD sensitive components or connectivity modes. We show that this framework (1) increases test-retest correspondence and (2) allows for better separation, in functional connectivity, of diagnostic groups both at the whole brain and individual resting state network level. Finally, we evaluate a posteriori the association between connectivity mode weights with longitudinal neurocognitive outcomes.

## 1   Introduction

Developing biomarkers for early detection of Alzheimer's disease (AD) is of critical importance as researchers believe clinical trial failures are in part due to testing of therapeutic agents too late in the disease [1]. The AD disconnection syndrome hypothesis [2] posits that AD spreads via propagation of dysfunctional signaling, indicating that functional connectivity (FC) biomarkers have potential for early detection. Despite this potential, known issues with high amounts of variability in acquisition and preprocessing of resting state fMRI, and ultimately low disease-related signal to noise ratio in FC [3], remain a critical barrier to incorporating FC as a clinical biomarker of AD. Recent work validated the utility of group level principal component analysis (PCA) to denoise FC by reconstructing subject level FC using PCs which optimized test-retest reliability through a measurement denominated differential identifiability [4]. Building on this work, we expand the utility of the framework to increase separation across diagnostic groups in the AD spectrum by reconstructing individual FC using AD sensitive PCs. We identify AD sensitive PCs using a novel diagnostic identifiability metric ($D$). We evaluate the proposed method with data from the Alzheimer's Disease Neuroimaging Initiative (ADNI2/GO) using group balanced, bootstrapped random sampling.

## 2 Methods

### 2.1 Subject Demographics

Of the original 200 ADNI2/GO individuals with resting state fMRI scans, subjects were excluded if they (1) had only extended resting state scans, (2) had no Amyloid status provided, (3) were cognitively impaired, but Amyloid-beta protein negative (Aβ-) negative, and/or had (4) over 30% of fMRI time points censored (see 2.2). The final sample included 82 individuals. Only Aβ positive (Aβ+) individuals were included in cognitively impaired groups to avoid confounding by non-AD neurodegenerative pathologies. Subjects were sorted into 5 diagnostic groups using criterion from ADNI2/GO and Aβ positivity: (1) normal controls ($CN_{Aβ-}$, n = 15), (2) pre-clinical AD ($CN_{Aβ+}$, n = 12), (3) early mild cognitive impairment ($EMCI_{Aβ+}$, n = 22), (4) late mild cognitive impairment ($LMCI_{Aβ+}$, n = 12), and (5) dementia ($AD_{Aβ+}$, n = 21). Aβ status was determined using either mean PET standard uptake value ratio cutoff (Florbetapir > 1.1, University of Berkley) or CSF Aβ levels [5]. Composite scores were calculated for visuospatial, memory, executive function, and language domains [6] from the ANDI2/GO battery. No demographic group effects were observed. All neurocognitive domain scores exhibited a significant group effect (Table1).

**Table 1.** Demographics and Neurocognitive Comparisons of Diagnostic Groups.

| Variable | $CN_{Aβ-}$ (n = 14) | $CN_{Aβ+}$ (n = 12) | $EMCI_{Aβ+}$ (n = 22) | $LMCI_{Aβ+}$ (n = 13) | $AD_{Aβ+}$ (n = 21) |
|---|---|---|---|---|---|
| Age (Years) (SD) | 74.2 (8.8) | 75.9 (7.0) | 72.6 (5.2) | 73.3 (6.1) | 73.5 (7.6) |
| Sex (% F) | 64.2 | 41.7 | 50 | 61.6 | 42.9 |
| Years of Education (SD) | 16.7 (2.3) | 15.8 (2.6) | 15.2 (2.6) | 16 (1.8) | 15.4 (2.6) |
| Visuospatial Domain Score (SD)** | 9.7 (0.61) | 9.3 (0.9) | 9.4 (0.9) | 83 (2.3) | 7.4 (2.1) |
| Language Domain Score (SD)** | 49.2 (4.2) | 48.8 (4.4) | 46.2 (5.8) | 43.1 (8.0) | 34.8 (9.6) |
| Memory Domain Score (SD)** | 125.4 (41.1) | 142 (34.5) | 104.9 (46.6) | 81.0 (36.7) | 34.2 (21.8) |
| Executive Function Domain Score (SD)** | 99.0 (26.8) | 117.6 (27.4) | 135.0 (48.6) | 166.3 (102.0) | 284.6 (101.0) |

** Significant group effect (Chi-square or ANOVA as appropriate, $\alpha = 0.05$)

### 2.2 fMRI Data Processing

MRI scans used for construction of FC matrices included T1-weighted MPRAGE scans and EPI fMRI scans from the initial visit in ADNI2/GO (www.adni-info.org for protocols). fMRI scans were processed in MATLAB using an FSL based pipeline following processing guidelines by Power et al [7] and described in detail in Amico et al. [8]. Subjects with over 30% of volumes censored due to motion were discarded to ensure data quality. For purposes of denoising FC matrices [4], processed fMRI time series were split into halves, representing "test" and "retest" sessions.

### 2.3 Test-Retest Identifiability and Construction and of Individual FC Matrices

For each subject, two FC matrices were created from the "test" and "retest" halves of the fMRI time-series. FC nodes were defined using a 286 region parcellation [9], as detailed in Amico et al. [8]. Functional connectivity matrices were derived by calculating the pairwise Pearson correlation coefficient ($r_{ij}$) between the mean fMRI time-series of all nodes. "Test" and "retest" FCs were de-noised by using group level PCA to maximize test-retest differential identifiability (Idiff) [4]. The "identifiability matrix" **I** was defined as the matrix of pairwise correlations (square, non-symmetric) between the subjects' $FC_{test}$ and $FC_{retest}$. The dimension of **I** is $N^2$ where N is the number of subjects in the cohort. Self-identifiability, ($I_{self}$, Eqn. 1), was defined to be the average of the main diagonal elements of **I**, consisting of correlations between $FC_{test}$ and $FC_{retest}$ from the same subjects. $I_{others}$ (Eqn. 2), was defined as average of the off-diagonal elements of matrix **I**, consisting of correlations between $FC_{test}$ and $FC_{retest}$ of different subjects. Differential identifiability ($I_{diff}$, Eqn. 3) was defined as the difference between $I_{self}$ and $I_{others}$.

$$I_{self} = \frac{1}{N} \sum_{i=j} I_{i,j} \quad (1)$$

$$I_{others} = \frac{1}{N} \sum_{i \neq j} I_{i,j} \quad (2)$$

$$I_{diff} = 100 * (I_{self} - I_{others}) \quad (3)$$

Group level PCA [10] was applied in the FC domain, on a data matrix ($\mathbf{Y_1}$) containing vectorized $FC_{test}$ and $FC_{retest}$ (upper triangular) from all subjects. PCs throughout this paper will be numbered in order of variance explained. The number of PCs estimated was constrained to 2*N, the rank of the data matrix $\mathbf{Y_1}$. Following decomposition, PCs were iteratively added in order of variance explained. Denoised $FC_{test}$ and $FC_{retest}$ matrices were reconstructed using the number of PCs (n) that maximized $I_{diff}$ (Eqn3), while maintaining a minimum $I_{others}$ value of 0.4, such that between-subject FC was neither overly correlated (loss of valid inter-subject variability) nor overly orthogonal (inter-subject variability dominated by noise). This was done because the ADNI2/GO fMRI data was noisier than data on which this method was previously implemented, as evidenced by a much lower original between-subject FC correlation ($I_{others}$ 0.22 ADNI vs. 0.4 Human Connectome Project rs-fMRI). Therefore, not setting a minimum threshold for $I_{others}$ led to the algorithm picking PCs that were "specialized" to specific subjects. The threshold 0.4 was specifically chosen because it reflected average $I_{others}$ values seen in FCs from previous data, on which this method was implemented [4].

Final, de-noised *FC* matrices were computed as the average of $FC_{test}$ and $FC_{retest}$. Nodes were assigned to 9 resting state subnetworks (RSN/RSNs), visual (VIS), somatomotor (SM), dorsal attention (DA), ventral attention (VA), limbic (L), fronto-parietal (FP), and default mode network (DMN) [11] with the additional subcortical (SUB) and cerebellar (CER).

## 2.4 Diagnostic Identifiability

With the goal of early detection in mind, we hypothesized that FC in non-dementia groups would become significantly less identifiable from FC in $AD_{A\beta+}$ with increased diagnostic proximity to $AD_{A\beta+}$. Figure 1 delineates the work flow for finding AD sensitive PCs using a novel diagnostic identifiability metric (*D*), which quantifies differentiability in connectivity between each non-dementia group (g) and $AD_{A\beta+}$ and is calculated from the correlation matrix (**I**) of $\mathbf{Y_2}$, containing final, de-noised FC from all subjects. $D_g$ was defined as the average correlation within a non-dementia group, *corr(g,g)*, minus the average correlation between that non-dementia group and $AD_{A\beta+}$, *corr(g, $AD_{A\beta+}$)*. *D*, rather than variance explained, was used to filter components, as it was hypothesized that early disease changes likely do not account for a large portion of between subject variance.

$$D_g = corr(g,g) - corr\big(g, AD_{A\beta+}\big) \quad (4)$$

Group level PCA was again performed on the matrix $\mathbf{Y_2}$. Here, the number of PCs was constrained to *n = 35* PCs, the rank of the $\mathbf{Y_2}$ matrix. $\mathbf{Y_2}$ was iteratively reconstructed using a subset of the n PCs, selected based on maximizing $D_g$. Starting with $PC_1$, $PC_{2...n}$ were iteratively added based on their influence on *average($D_g$)*. At each iteration, the $PC_{j*}$ which most improved *average($D_g$)* upon its addition to previously selected PCs, was selected. To avoid results driven by a subset of the population or by differences in sample sizes between groups, the cohort was randomly sampled 30 times, following total cohort PCA, in a group balanced fashion ($n_{sample}$ = 50; $n_g$ = 10). The number of bootstraps was chosen to allow adequate estimation of the $D_g$ distribution while keeping run-time of the algorithm, reasonable. Bootstrapped distributions of $D_g$ were generated for each number of PCs. The number of PCs (n*) which maximized average($D_g$) was found (Eqn. 5). AD sensitive PCs were defined as those which appeared within the n* most influential PCs with the greatest frequency across samples. Final FC matrices were re-constructed using only AD sensitive PCs.

$$n^* = n, at\ argmax_n \big(\frac{1}{g}\sum_g D_g(\text{n})\big) \quad (5)$$

Additionally, $D_g$ curves were estimated and disease sensitive PCs were identified for the 9 RSNs individually, by calculating $I_{RSN}$ using the subset of connections where at least one of the nodes in the connection was part of the RSN.

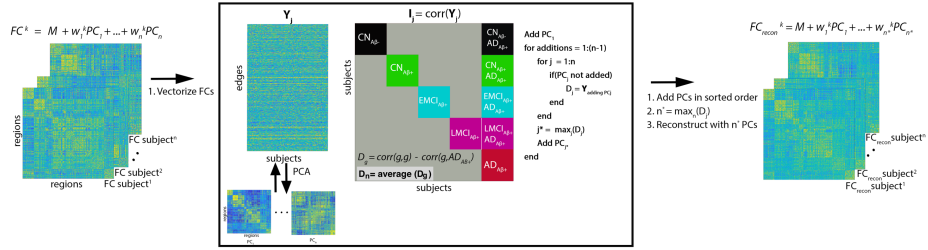

**Fig. 1.** Diagnostic Identifiability workflow.

### 2.5 Statistical Validation and Association with Neurocognitive Outcomes

Due to the small number of bootstraps, differences between $D_g$ distributions were assessed at n* PCs by checking if the median of one distribution was an outlier relative to a reference distribution using non-parametric confidence intervals defined with the median and interquartile range (IQR). First, $D_g$ distributions from each RSN were compared to those from WB. Next, WB and RSN $D_g$ distributions were compared to a corresponding *null* model. *Null* models for the WB and each RSN were constructed by randomly permuting diagnostic group membership among individuals selected at each bootstrap, such that $D_g$ for the *null* model represented identifiability of a random heterogeneous group from a random heterogeneous reference group. Finally, individual $D$ values ($D_i$) were calculated for each subject using FC reconstructed with the n* PCs. ANOVA ($\alpha < 0.05$) with follow up pairwise tests, was performed on WB $D_i$ distributions to test for a group effect. Stepwise regressions (F-test, $\alpha = 0.05$), starting with gender, age and education, were be used to test for associations between the n* PC weights and longitudinal changes in neurocognitive outcomes (0, 1, 2 years post imaging).

## 3 Results

### 3.1 Test-Retest Identifiability

Figure 2 details the results of denoising FC using differential test-retest differential identifiability. An optimal reconstruction based on the first n = 35 PCs (in decreasing order of explained variance) was chosen (Figure 2A). $I_{self}$ increased from 0.52 to 0.92 (Figure2A-B) while $I_{others}$ increased from 0.20 to 0.40 (Figure 2A-B). $I_{diff}$ increased from 38% to 57% (Figure2A-B).

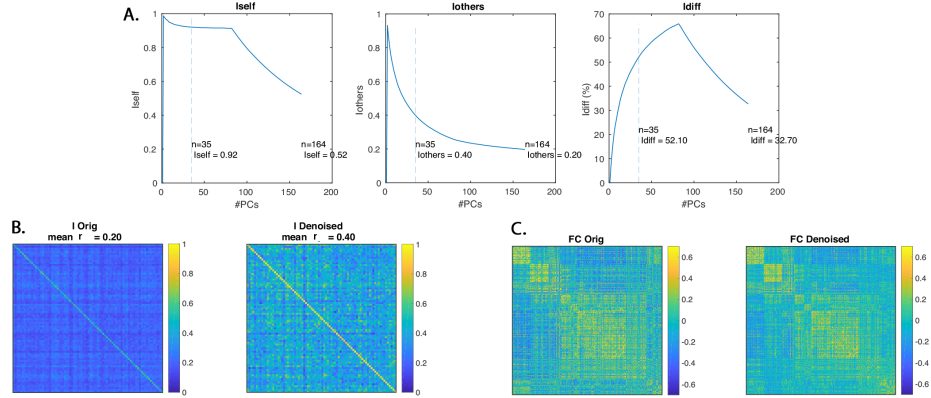

**Fig. 2.** (A) $I_{self}$, $I_{others}$, and $I_{diff}$ across the range of # PCs. (B) I matrices for original and denoised FC matrices. (C) Example original FC matrix versus denoised FC matrix.

### 3.2 Diagnostic Identifiability

WB average($D_g$) peaked at n* = 11 components which explained 58.82% of the variance in the denoised FC data (Figure 3A, Table2). At n* PCs, $LMCI_{Aß+}$ was the only group who that did not exhibit significantly increased $D_g$ from the *null* model. At n* components, $D_i$ distributions exhibited a significant group effect. $D_i$ decreased with diagnostic proximity to $AD_{Aß+}$ (Figure 3B). Between-subject correlation in FC increased from 0.41 to 0.71 after reconstruction with n* PCs (Figure 3B). Of the 9 RSNs, the L network exhibited significantly greater $D_{RSN}$ as compared to WB (Table2). Like WB, $LMCI_{Aß+}$ was the only group that did not exhibit significantly greater RSN $D_g$ than the null model, with the exceptions of SM where $EMCI_{Aß+}$ was additionally not significantly different from the *null* model and L where all non-dementia groups exhibited greater $D_g$ than the *null* model (Table2). Eight of eleven PCs were identified as disease sensitive in all 9 RSNs and WB (**Table 2**).

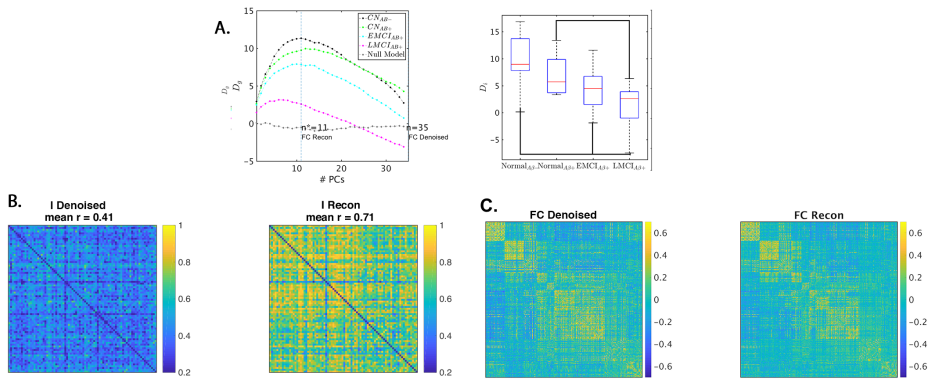

Fig. 3. (A-left) Whole brain $D_g$ across all possible number of PCs. (A-right) Individual $D_i$ values at n* = 11 PCs. Distributions showing significant differences (t-test, $p < 0.05$) are delineated using lines. (B) Denoised I matrix (n = 35 PCs) versus I matrix reconstructed using disease sensitive PCs (n* = 11 PCs). (C) Example denoised FC matrix versus FC matrix reconstructed using disease sensitive PCs.

Table 2. Diagnostic Identifiability Summary.

| RSN | $CN_{A\beta-}$ | $CN_{A\beta+}$ | $EMCI_{A\beta+}$ | $LMCI_{A\beta+}$ | Mean | n | Var (%) |
|---|---|---|---|---|---|---|---|
| WB | 11.35** | 9.75** | 7.85** | 2.60 | 7.89 | 11 | 58.82 |
| VIS | 13.21** | 10.33** | 8.11** | 2.75 | 8.60 | 10 | 57.23 |
| SM | 9.82** | 12.96** | 7.30 | 4.43 | 8.62 | 10 | 57.26 |
| DA | 12.16** | 11.24** | 8.09** | 3.37 | 8.71 | 13 | 62.19 |
| VA | 10.74** | 11.65** | 7.96** | 2.33 | 8.17 | 10 | 57.26 |
| **L** | **17.18**** | **13.76**** | **11.97**** | **6.28**** | **12.30** | **8** | **54.50** |
| FP | 12.07** | 10.25** | 9.34** | 2.70 | 8.59 | 11 | 58.82 |
| DMN | 12.09** | 10.20** | 8.39** | 2.84 | 8.38 | 11 | 58.82 |
| SUB | 14.17** | 11.66** | 9.93** | 4.33 | 10.02 | 9 | 55.78 |
| CER | 13.29** | 12.85** | 9.72** | 5.67 | 10.38 | 10 | 57.26 |

**Median outside CI null model, **Median outside CI WB mean($D_g$)**

Four PCs exhibited significant associations with various neurocognitive domain scores (Table 3). Visuospatial domain scores were associated with PC 17 at 1 year post imaging and PC 9 at 2 years post imaging. Memory domain scores were associated with PC 32 at 1 year post imaging and PC 7 at 2 years post imaging. Language domain scores were associated with PC 23 at 0 year post imaging and PC 7 at 1 years post imaging. Finally, PC 17 was associated with executive domain scores at 1 and 2 years post imaging.

Table 3. Associations of n* PC weights with neurocognitive composite domain scores. Stepwise regressions (F-test, $\alpha < 0.05$) were used to assess the relationship of neurocognitive composite domain scores with PC weights, with age, gender, and education starting in the base model; $p$ values are reported for the whole model, adjusted-$R^2$ is reported for the model.

| Time Points | Visuospatial | | | Memory | | | Language | | | Executive | | |
|---|---|---|---|---|---|---|---|---|---|---|---|---|
| | PC | $p$ | $R^2$ | PC | $p$ | $R^2$ | PC | $p$ | $R^2$ | PC | $p$ | $R^2$ |
| 0 | - | - | - | - | - | - | 23 | 0.040 | 0.19 | - | - | - |
| 1 | 17 | 0.001 | 0.53 | 32 | 0.032 | 0.31 | 7 | 0.025 | 0.31 | 17 | 0.004 | 0.48 |
| 2 | 9 | 0.020 | 0.46 | 7 | 0.044 | 0.20 | - | - | - | 17 | 0.013 | 0.36 |

## 4 Limitations, Future Work, and Conclusions

We present here a two stage PCA based framework to improve the detection of AD signatures in whole-brain functional connectivity. We first use recently proposed test-retest differential identifiability to denoise subject-level functional connectomes and consequently reduce dimensionality of functional connectomes. We subsequently introduce and validate the concept of PCA based differential diagnostic identifiability to increase AD signal to background in functional connectivity. The result of a significant diagnostic group effect in diagnostic differential identifiability shows that FC contains AD signature, even at early stages of disease. The finding of increased diagnostic identifiability in Limbic regions, known to be associated with memory processes and known to be affected in AD, further validates this finding. Finally, we show that PC weights from AD sensitive principal components are correlated to longitudinal neurocognitive outcomes. In addition to the work presented here, we plan to delve further into the meaning of the PCs themselves. AD sensitive PCs did not appear to be specific to individual RSNs, as the same PCs were consistently AD sensitive across RSNs. Furthermore, several PCs were associated with multiple neurocognitive domains. Therefore, AD sensitive PCs may characterize global brain changes related to AD. However, spatial representation of PCs and relationship of PCs with network properties need to be explored to further assess this. Finally, to further validate these promising results, this methodology needs to be applied to a larger cohort. With ADNI3 data becoming available (~300 subjects already scanned), on which all subjects underwent resting state fMRI, we will be able to further validate findings and further improve identification and characterization of AD sensitive PCs based on whole brain functional connectomes. This dual decomposition/reconstruction framework makes forward progress in exploiting the clinical potential of functional connectivity based biomarkers.

## 5   References

1. Sperling, R.A., J. Karlawish, and K.A. Johnson, *Preclinical Alzheimer disease-the challenges ahead.* Nat Rev Neurol, 2013. **9**(1): p. 54-8.
2. Brier, M.R., J.B. Thomas, and B.M. Ances, *Network dysfunction in Alzheimer's disease: refining the disconnection hypothesis.* Brain Connect, 2014. **4**(5): p. 299-311.
3. Braun, U., et al., *Test-retest reliability of resting-state connectivity network characteristics using fMRI and graph theoretical measures.* Neuroimage, 2012. **59**(2): p. 1404-12.
4. Amico, E. and J. Goni, *The quest for identifiability in human functional connectomes.* Sci Rep, 2018. **8**(1): p. 8254.
5. Shaw, L.M., et al., *Cerebrospinal fluid biomarker signature in Alzheimer's disease neuroimaging initiative subjects.* Ann Neurol, 2009. **65**(4): p. 403-13.
6. Wilhalme, H., et al., *A comparison of theoretical and statistically derived indices for predicting cognitive decline.* Alzheimers Dement (Amst), 2017. **6**: p. 171-181.
7. Power, J.D., et al., *Spurious but systematic correlations in functional connectivity MRI networks arise from subject motion.* Neuroimage, 2012. **59**(3): p. 2142-54.
8. Amico, E., et al., *Mapping the functional connectome traits of levels of consciousness.* Neuroimage, 2017. **148**: p. 201-211.
9. Shen, X., et al., *Groupwise whole-brain parcellation from resting-state fMRI data for network node identification.* Neuroimage, 2013. **82**: p. 403-15.
10. Hotelling, H., *Analysis of complex variables into principal components.* Journal of Educational Psychology, 1933. **24**: p. 417-441.
11. Yeo, B.T.T., et al., *Estimates of segregation and overlap of functional connectivity networks in the human cerebral cortex.* NeuroImage, 2014. **88**: p. 212-227.